\documentclass[12pt,preprint]{aastex}

\usepackage{epsfig}

\newcommand{\mdot}{\ensuremath{\dot{M}}}
\newcommand{\lsun}{\ensuremath{\, {\rm L}_\odot}}
\newcommand{\msun}{\ensuremath{\, {\rm M}_\odot}}
\newcommand{\n}{\ensuremath{\mem{n}}}

\newcommand{\p}{\ensuremath{\mem{p}}}

\newcommand{\hevi}{\ensuremath{^{4}\mem{He}}}

\newcommand{\cdr}{\ensuremath{^{13}\mem{C}}}
\newcommand{\czw}{\ensuremath{^{12}\mem{C}}}

\newcommand{\ndr}{\ensuremath{^{13}\mem{N}}}
\newcommand{\nvi}{\ensuremath{^{14}\mem{N}}}

\newcommand{\ose}{\ensuremath{^{16}\mem{O}}}
\newcommand{\osi}{\ensuremath{^{17}\mem{O}}}
\newcommand{\oac}{\ensuremath{^{18}\mem{O}}}

\newcommand{\nezwa}{\ensuremath{^{20}\mem{Ne}}}

\newcommand{\nezw}{\ensuremath{^{22}\mem{Ne}}}

\newcommand{\nadr}{\ensuremath{^{23}\mem{Na}}}
\newcommand{\mgvi}{\ensuremath{^{24}\mem{Mg}}}
\newcommand{\mgfu}{\ensuremath{^{25}\mem{Mg}}}
\newcommand{\mgse}{\ensuremath{^{26}\mem{Mg}}}

\newcommand{\alse}{\ensuremath{^{26}\mem{Al}}}

\newcommand{\fese}{\ensuremath{^{56}\mem{Fe}}}

\newcommand{\zrse}{\ensuremath{^{96}\mem{Zr}}}
\newcommand{\zrvi}{\ensuremath{^{94}\mem{Zr}}}

\newcommand{\mem}[1]{\ensuremath{\mathrm{ #1}}}
\newcommand{\spr}{\mbox{$s$-process}}
\newcommand{\kelv}{\ensuremath{\,\mathrm K}}
\newcommand{\abb}[1]{Fig.\,\ref{#1}}
\newcommand{\kap}[1]{\S\,\ref{#1}}
\newcommand{\jahre}{\ensuremath{\, \mathrm{yr}}}

\newcommand{\tab}[1]{Table\,\ref{#1}}
 
\newcommand{\mhe}{\ensuremath{M_{\rm He}}}
\newcommand{\mh}{\ensuremath{M_{\rm H}}}        

\slugcomment{Version 1.3, \today}

\usepackage{natbib}

\shorttitle{Intermediate mass stars at low Z}
\shortauthors{F. Herwig}

\received{}
\begin{document}

\title{Dredge-up and envelope burning in intermediate 
mass giants of very low metallicity} 
\author{Falk Herwig}
\affil{Department of Physics and Astronomy, University of Victoria,
  3800 Finnerty Rd, Victoria, BC, V8P 1A1} 
\affil{Los Alamos National Laboratories, Los Alamos, NM 87544} \email{fherwig@lanl.gov}

\begin{abstract}
The evolution of intermediate mass stars at very low
metallicity  during their final
thermal pulse asymptotic 
giant branch phase is studied in detail. As representative examples
models 
with initial masses of $4\msun$ and $5\msun$ with a metallicity of
$Z=0.0001$ ($\mem{[Fe/H]}\sim-2.3$) are discussed.  The 1D stellar
structure and evolution model includes time- and depth dependent
overshooting motivated by hydrodynamical simulations, as well as a
full nuclear network and time-dependent mixing. Particular
attention is given to high time and 
space resolution to avoid numerical artefacts related to third dredge-up and
hot-bottom burning predictions. The model calculations predict very
efficient third 
dredge-up which  mixes the envelope with the entire intershell
layer or a large fraction thereof, and in some cases penetrates into
the C/O 
core below the He-shell. In all cases primary oxygen is mixed into
the envelope. The models predict efficient envelope burning during the
interpulse phase. Depending on the envelope burning temperature, oxygen
is destroyed to varying degrees. The combined effect of dredge-up and
envelope burning 
does not lead to any significant oxygen depletion in any of the cases
considered in this study. The large dredge-up efficiency in our model is
closely related to the particular properties of the H-shell during the
dredge-up phase in low-metallicity very metal poor stars,
which is followed here  over many
thermal pulses. During the dredge-up phase, the temperature just below
the convective boundary is large enough for protons to burn vigorously
when they are brought into the C-rich environment below the convection
boundary by the time- and depth dependent overshooting. H-burning
luminosities of 
$10^5$ to $\sim 2\times 10^6\lsun$ are generated. C and, to lesser degree O, is
transformed into N in this dredge-up overshooting
layer and enters the envelope. The global effect on the CNO abundance
is similar to that of hot bottom burning. If the overshoot efficiency is
larger, then dredge-up H-burning causes a further increase in the
dredge-up efficiency. After some thermal pulses the dredge-up proceeds
through the He-shell  and into the CO core beneath. Then neutrons may
not be released from 
\cdr\ in radiative conditions during the interpulse phase because of
the scarcity of $\alpha$-particles for the $\cdr(\alpha,\n)\ose$
reactions. Conditions for the \spr\ are discussed
qualitatively. The abundance evolution of H, He, C, N, O and Na is
described. Finally, the model predictions for sodium and oxygen
are compared with observed abundances. The notion
that  massive  AGB stars are the origin of the O-Na abundance
anti-correlation in globular cluster giants is not consistent with the
model predictions of this study. The abundance of the C-rich
extremely metal poor binaries LP\,625-44, CS\,29497-030 and HE\,0024-2523 is
discussed. 
\end{abstract}


\section{Introduction}
\label{sec:intro}
During their final evolutionary phase intermediate mass stars (IMS) proceed
through the  
thermal pulse Asymptotic Giant Branch (TP-AGB) stage
\citep[e.g.][]{iben:83b,lattanzio:99,herwigIAU209_01}. Hydrogen and 
helium is processed in distinct nuclear shells around the increasingly
electron-degenerate CO-core. Periodically the He-shell becomes
unstable due to a 
combination of the well-known thin-shell instability and partial
degeneracy. The thin-shell instability is a common feature in nuclear
shells around degenerate cores, and can be found in accreting white
dwarfs (novae), in He-core white dwarfs as well as in accreting
neutron stars (X-ray bursts). In   
AGB stars the He-shell flash causes a complicated series of mixing
events, including a pulse-driven convection zone (PDCZ) which
comprises the entire intershell layer. These mixing events eventually
lead to a significant envelope abundance enrichment 
with material processed by either or both the He-shell and the
H-shell. 

In the past most AGB stellar evolution calculations have been carried
out for the solar metallicity, moderate metal deficiency, or
zero-metallicity \citep[e.g.][]{karakas:02a,siess:02}. However,
there are a number of reasons to study the evolution of AGB stars
at very low non-zero metallicity ($0.0<Z\leq0.0001$). 
For example, recent observations of \nvi\
in metal poor stars and, in particular, in the Damped Lyman-$\alpha$ absorbers
\citep{pettini:02,prochaska:02} have sparked a debate on whether the
results are to be interpreted as a primary origin of \nvi\ in
IMS or as a result of a peculiar (top-heavy)
initial mass function at very 
low metallicity. 

Detailed abundance studies of extremely metal poor (EMP) carbon-rich stars
are now reported in increasing numbers
\cite[e.g.][]{lucatello:02,aoki:02a}. Many of these carbon rich EMP stars
are nitrogen-rich as well, and often -- when it has been determined --
these objects show elevated oxygen, sodium and \spr\ element
abundances.  A similar pattern with respect to C, N, and the
\spr\ elements has 
been reported in mildly metal-poor  post-AGB stars
\citep{reddy:02,vanwinckel:00}. However, since these stars 
are only moderately metal poor, it is conceivable that
this nitrogen abundance is of secondary origin, and a result of the
first dredge-up (DUP). This possibility can be excluded for nitrogen
in most C-rich EMP stars, where the N-abundance is much larger than
the sum of the initially present C, N, and O abundances. In those cases
where the binarity of a C-rich EMP star is established, the abundance of
nitrogen and other elements may very well be the result of mass
transfer from an AGB 
companion that now has evolved into a white dwarf. 

Intermediate-mass stars may contribute to the abundance anomalies in
globular cluster 
stars in the self-pollution scenario
\citep{cottrell:81,denissenkov:97}. In particular, the observation of 
abundance anti-correlations in turn-off main-sequence stars
\citep{gratton:01} has 
made it necessary to consider an external source for the abundance
anomalies. \citet{ventura:02} have proposed that the
oxygen-sodium anti-correlation is due to the pollution of globular cluster
members with material blown off  TP-AGB stars. However, they
have not explicitly included the Ne-Na and the 
Mg-Al H-burning cycle in their calculations of TP-AGB stars with $Z \ge
0.0004$. These reactions have been taken into account by
\citet{denissenkov:03a}, whose short study is partly
based on models presented here in more detail. Finally, it should be
noted that the TP-AGB phase of IMS is the dominant source of dust in early
galaxies of redshifts up to 5 \citep{morgan:03}.

IMS during their  TP-AGB evolution  change their surface composition by two
processes, the third dredge-up (DUP) and envelope burning between thermal
pulses (which is also known as hot bottom burning, HBB). Low-mass 
stellar models of Pop~I metallicity with a  
combination  of convective overshooting and high spatial 
and time resolution can reproduce formation of low luminosity carbon
stars needed to explain the carbon star luminosity function in the
Magellanic Clouds \citep{herwig:97,herwig:99a,mowlavi:99}.
Other consistency checks concerning the efficiency of overshooting at
the various boundaries involve the effects of  \spr\
nucleosynthesis \citep{lugaro:02a,herwig:02a}. 
The picture that has
emerged from these studies of low-mass AGB stars that do not
experience HBB is summarized in \kap{sec:ov}.
 In \kap{sec:code} the stellar evolution
code and the model set is described. The results
are presented in \kap{sec:results}, and conclusions are given in
\kap{sec:concl}. 


\section{Treatment of convective boundaries}
\label{sec:ov}

Hydrodynamic time- and depth-dependent overshooting may
be present at all convective boundaries. An extended discussion of the
motivation and background, as well as the technique of the exponential
overshoot
implementation used in this study can be found in \citet{herwig:99a} and
\citet{herwig:02a}. The overshooting
efficiency in the deep interior cannot be determined from
hydrodynamic models ab initio due to the large range of scales
involved. In order to construct AGB models with
some predictive power, one is therefore forced to constrain the
overshooting efficiency semi-empirically. At the bottom of the
envelope of TP-AGB stars, exponential overshooting 
increases the DUP depth at almost any non-zero efficiency. With
an efficiency $f\approx 0.15$, the resultant \cdr\ pocket that forms
during the final period of the third DUP is massive enough to
reproduce the observed \spr\ overabundances
\citep{goriely:00,lugaro:02a,herwig:02a}. Models with overshooting at 
the bottom of the  PDCZ  predict even more efficient third DUP
as well as a larger carbon and oxygen abundance in the intershell
zone, compared with models that do not include overshooting at this
convective boundary. The larger carbon and, in particular, the larger
oxygen abundance in the intershell is
in good agreement with those observed in H-deficient central stars of
planetary nebulae \citep[{of spectral type PG1159 and
[WC]-CSPN\footnote{Central stars of planetary nebulae with Wolf-Rayet
  type spectrum.},}][]{koesterke:97b,dreizler:96} in conjunction with the 
latest models on the evolutionary origin of these stars
\citep{herwig:99c,herwig:01a}. These models 
predict that the  PG1159 and [WC]-CSPN stars show on their surfaces the
abundance distribution of the the intershell layer of the progenitor
AGB star. In this context it is interesting to note that  recent
FUSE spectroscopic observations have established a significant
Fe-deficiency in  some hot H-deficient central stars \citep{miksa:02}. Further
observations may substantiate the hypothesis that this Fe-depletion is
a result of the \spr\ in the intershell during the progenitor AGB
evolution. This would  
support the claim that hot H-deficient central stars of PN display
directly the intershell material of their AGB progenitors.

However, in the \spr\ models, the larger carbon intershell abundance
in models with overshooting leads to a 
larger neutron 
exposure in the \spr\ layer during the interpulse phase
\citep{lugaro:02a}. In these 
models the \spr\ abundance distribution is too top-heavy, or in other
words the [hs/ls] index is larger than observed. This could be
compensated by some weak mixing of the \spr\ layer during the
interpulse phase  \citep[e.g.\ as a result of slow rotation, ][]{herwig:02a}. 

Models with exponential overshooting at the bottom of the PDCZ predict larger
temperatures in the second \spr\ production location, fuelled by the
\nezw\ neutron source. Some branchings like the
\zrse/\zrvi\ ratio are sensitively dependent on the temperature
at the bottom of the PDCZ. At the same time, such isotopic ratios are
experimentally known to high precision from mainstream pre-solar SiC
grains, steming from low-mass AGB
stars \citep{zinner:98,nicolussi:98b}. This puts perhaps the 
most stringent constraint on the maximum possible overshooting
efficiency. Although a more detailed study in this respect is still
pending, preliminary calculations by \citet{lugaro:02a} indicate that
low mass models with $f_\mem{PDCZ} = 0.008$ may be in agreement with
the temperature-dependent branchings, and at the same time retain the
enhanced oxygen intershell abundance indicated by the PG1159 and
[WC]-CSPN stars. 

\section{Stellar evolution code and model calculations}
\label{sec:code}
The models in this study were computed with the full 1D stellar
evolution code of
\citet{herwig:99a}, including the exponential overshooting model
described there. An earlier version of this code has been used by
\citet{bloecker:95a} to study the evolution of Pop.\,I AGB stars as a
function of initial mass. As in most other AGB models the
Schwarzschild critierion for convection is used. The
stabilizing effect of a $\mu$-gradient as reflected by the Ledoux
criterion is thus neglected. Using the Ledoux criterion in AGB stars
would make it even more difficult to obtain the third dredge-up. In
addition, the exponential  overshooting included here,
would in this situation likely dominate mixing through semiconvection. 
In order to meet the high numerical resolution requirements of third DUP
models, the adaptive time step and grid size routine
was improved. It can now deal with the hot DUP burning
episodes described below. In particular, it is ensured that the time-step
and the grid resolution correspond to each other in the region near
the bottom of the convective envelope during the DUP phase. In
this situation, the Lagrangian advance rate of the bottom of the convective
envelope $\dot{m}_\mem{CE}$ implies a relation between the grid
resolution and the time 
step. Numerical experiments indicate an appropriate Lagrangian
scale in the third DUP overshooting zone to be $l_\mem{m}=10^{-6}\msun$,
which should be well sampled in space and time. This means that
the typical grid spacing in the DUP zone is $<10^{-7}\msun$
and time steps should not be larger than $l_\mem{m}/\dot{m}_\mem{CE}$.
In practice, this means that a single third DUP episode  of
several $10^{-3}\msun$ requires  computing several thousand models. In
the case where nuclear burning takes place during the DUP phase
-- as described below -- the time step is further limited by the
fluctuations of the hydrogen-burning luminosity $L_\mem{H}$. 

The opacities are from \citet[][OPAL]{iglesias:96} 
supplemented with low temperature opacities by
\citet{alexander:94}. The mixing-length parameter has been set to
$\alpha_\mem{MLT}=1.7$, as calibrated by fitting a solar model. 
The nuclear network equations and the
time-dependent mixing  equations are solved simultaneously, using a
fully implicit, and iterative scheme. For this study an automatic
sub-time step algorithm has been implemented into the Newton-Raphson
scheme. If the convergence criteria are not meet within five to eight
iterations the time step is replaced by 10 times smaller sub-time
step, followed by a second sub-time step that takes care of the
remaining $90\%$ of the original time step. This procedure can be
applied repeatedly,  and guarantees that a solution with the
pre-defined precision can be found even in fairly violent burning
and mixing events, as encountered in massive AGB models of very low
metallicity. All relevant charged particle reactions are included up to and
including the Mg-Al cycle, as well as neutron captures. Reaction rates
are mainly based on \citet{caughlan:88}. Updates from
\citet{eleid:95} and from \citet{angulo:99} for the important CN-cycle
p-capture rates have been considered.

The benchmark model sequence (E79-D4, see \tab{tab:ovdep}) has an initial
mass of  $5\msun$ and  a metallicity of $Z=0.0001$
([Fe/H]\,$\approx\log(Z/Z_\odot)$\,=\,$-2.3$). It  is evolved from
the pre-main sequence through all evolutionary phases up to the
AGB. 
For the initial helium mass-fraction abundance we adopt $Y=0.23025$  according to
\begin{displaymath}
Y=Y_\odot + \frac{\Delta Y}{\Delta Z} \times (Z-Z_\odot) ,
\end{displaymath}
where $Y_\odot=0.28$ is the value adopted for the solar initial helium
abundance \citep{grevesse:93,grevesse:98}, and $Z_\odot=0.02$ has been
adopted for the solar metallicity. $\Delta Y 
/ \Delta Z $ is the ratio of fresh helium to metals supplied to the interstellar
medium by stars. The value
of 2.5 used in this study is consistent with many observational
constraints \citep[e.g.][and references therein]{pagel:98}. 

Many stars of low metallicity may, in fact, form from gas clouds that
have a non-solar abundance distribution. For example, \citet{carney:96}
concluded in his 
review of  observational data that globular cluster stars  with
$\mem{[Fe/H]}\sim -0.6$ are overabundant in the $\alpha$-elements by a
factor of about 2. However, some halo stars have
been reported with solar-scaled $\alpha$-element abundances
\citep{nissen:97}. 
For this study we have adopted a scaled solar
metal distribution according to \citet{grevesse:93} and 
\citet{anders:89}. Most of the results discussed here are related to
the production and  processing of primary C and O, which dominate the
surface abudance evolution once DUP starts.

Mass loss has not been considered for the $5\msun$ cases, on which
most of this investigation is based. This choice
was motivated by the fact that model tracks with mass loss show less
DUP \citep{karakas:02a}. Besides, the amount of mass
loss that occurs in very 
low metallicity stars is uncertain. Due to the lack of mass loss the
calculations do not have a natural end due to the loss of the stellar
envelope. At least one sequence is followed through 15 TPs which is
sufficient to reach the asymptotic regime in which changes of many
quantities are smaller  and more regular than during the initial TP-AGB phase.

Up to the first TP on the AGB, exponential  overshooting with $f=0.016$
is applied at all convective
boundaries. This is the same value as 
used by \citet{herwig:99a}, and it was
originally obtained by calibrating core-overshooting to reproduce the
observed width of the 
main-sequence.   For the TP-AGB benchmark
case (sequence E79-D4) a non-zero but negligible overshooting 
efficiency\footnote{This small amount of overshooting has very little
  effect on the He-flash peak luminosities and intershell abundances,
  but behaves numerically more benign with the simultaneous solution
  of time-dependent mixing and nuclear burning.} of
$f_\mem{PDCZ}=0.002$ at the bottom of the 
PDCZ and the  
previously used value of $f_\mem{CE}=0.016$ at the bottom of the 
convective envelope are applied at all times -- during the HBB
interpulse phase as well as during the DUP phase. In our opinion, this
choice 
represents a minimum overshooting situation. AGB models with such
overshooting parameters would not generate enough \cdr\ for the \spr\
production during the intershell, and they would not be able to
reproduce the observed abundances of PG1159 stars and [WC]-CSPN
(\kap{sec:ov}).  For this reason, the possibility of larger overshooting
efficiency is explored by a number of test computations
(\tab{tab:ovdep}, \kap{sec:depov}). 
Rotation and the possible effects of magnetic fields are not included.

\section{Results}
\label{sec:results}

\subsection{Pre-AGB evolution}
The 5\msun\ pre-main sequence model is evolved through all phases up
to the giant stage. The evolution track is in good
agreement with  calculations by \citet{girardi:96}
(\abb{fig:hrd-preAG}). The small   
remaining differences can be attributed to differences in overshooting
efficiency, mixing-length parameter, and other details of the physical
inputs and numerical techniques. The core H- and He-burning times -- in
particular the sum of both -- are in reasonable agreement as
well. This model has 
$\tau_\mem{H}=8.64\times10^7\jahre$ and $\tau_\mem{He}=1.21\times10^7$ compared to $\tau_\mem{H}=9.00\times10^7\jahre$ and
$\tau_\mem{He}=9.13\times10^6\jahre$ for the model by
\citet{girardi:96}. 
Initially the mass of the main-sequence core convection zone is
$1.33\msun$, and decreases to $\sim0.45\msun$ shortly before H-burning
in the core ceases. The maximum extent of the convective He-burning
core is $0.55\msun$ and one major breathing pulse has been found. At
the end of the core He-burning phase, the H-shell is located at
$m_\mem{r}=1.47\msun$. The second DUP starts $4.25\times10^6\jahre$
after the end of the He-core burning  when the bottom of the
convective envelope reaches into the H-free core. This mixing episode
ends after $63000\jahre$ when the bottom of the envelope has reached
$m_\mem{r}=0.9615\msun$. 

Intermediate-mass stars at this low metallicity do not have a first giant
branch phase \citep{girardi:96} and thus the first envelope
abundance alteration occurs as a result of the second
DUP. Material processed mainly by
H-shell burning is brought to the surface. \hevi\ is enriched by about
$50\%$ to $X(\hevi)=0.335$ and hydrogen is reduced accordingly to
$X(\mem{H})=0.65$. Compared to the  very efficient 
production of \nvi\ 
by the combined action of HBB and the third DUP later during the
TP-AGB evolution, the second DUP of
\nvi\ with an enhancement by $\sim 50\%$ can be considered
unimportant. 
 \nadr\ is enhanced by
$0.75\mem{dex}$ from the conversion of the initial abundance of \nezw\
and some \nezwa\ (see also \kap{sec:oxysod}). \ose\  is depleted by
$0.1\mem{dex}$ due to partial ON cycling. Magnesium isotopes are changed by less than $0.05$,
$0.1$ and $0.01\mem{dex}$ for mass numbers $24$, $25$ and $26$ in the
second DUP. The first He-shell flash occurs  $12200\jahre$ after the
end of the second DUP. At this time the  H-free core has a 
mass of $0.9630\msun$.

\subsection{Envelope burning and third dredge-up}
\label{sec:pulsetdup}
\begin{figure}
\plotone{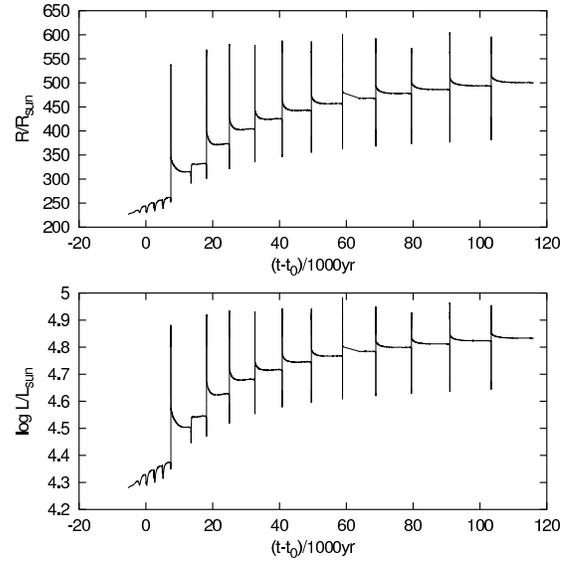}
\figcaption{ \label{fig:rad-Lsurf} 
Evolution of the stellar radius (top panel) and the stellar luminosity
(bottom panel) for the same sequence as shown in \abb{fig:mh1}.
}\end{figure}

The entire chemical enrichment is strongly dependent on the occurrence
of the third DUP. The 5\msun\ TP-AGB model sequence  shows
efficient third DUP after the fourth thermal pulse (\abb{fig:mh1}). The DUP
mass is almost $\Delta M_\mem{DUP}=3\times10^{-3}\msun$, comparable 
to the DUP in a low-mass AGB star with
$\lambda=0.5$. This DUP parameter is defined as
$\lambda=\Delta M_\mem{H}/\Delta M_\mem{DUP}$, where $\Delta
M_\mem{H}$ is the core mass growth due to H-burning during the
interpulse phase.  All following thermal pulses
have a very efficient third DUP as well. 
The fifth thermal pulse is an exception because the time-step
control algorithm failed. As a result, the DUP efficiency is
too small (\abb{fig:mh1}). For all subsequent TPs the DUP parameter always
exceeds unity, and the H-free core \mh\ effectively shrinks slightly. This
is also true for the He-free core \mhe. The evolution of the He- and
H-burning luminosity, as well as the stellar radius and the stellar
luminosity, of the $5\msun$ model is shown in \abb{fig:lumi} and
\ref{fig:rad-Lsurf}. Note how the surface properties as well
as the evolution of the shells, change at the onset of deep
dredge-up. The evolution of shell luminosities and surface parameters
differs strongly from the situation in more metal-rich and/or less
massive models \citep{lattanzio:92,bloecker:95a,herwig:99a}.

A more detailed presentation of the sixth thermal pulse is given in
\abb{fig:t-mh1-tp6}. The evolution of the bottom of the envelope
convection zone is very smooth. In various test sequences the time
evolution of this convective boundary showed discontinuous step-like
advances into the core. In these tests the numerical resolution was
insufficient. Such features cannot be found during the
DUP after the sixth or any subsequent thermal pulse of the benchmark
5\msun\ sequence. 
In this model sequence the third DUP, in particular, with 
respect to the Lagrange location of the He-burning shell (\mhe) and the
PDCZ, is very deep. The envelope convection not only engulfs the entire 
interpulse zone, but it even penetrates into the He-free core. This third
DUP behavior is not present in more metal-rich massive AGB
models or in low-mass AGB models. A decreasing \mhe\ has previously
been reported by \citet[][Table 1]{herwig:99a} for the TP-AGB
evolution of a $M_\mem{ini}=4\msun$ sequence with $Z=0.02$. However,
for that sequence, efficient overshooting ($f_\mem{PDCZ}=0.016$)  at
the bottom of the PDCZ has been assumed, which leads to a deep
penetration of the PDCZ into the C/O core beneath the He-shell. This
is not the case in the present model, which has been computed with negligible
overshooting at the He-shell flash boundary
($f_\mem{PDCZ}=0.002$). The bottom of the PDCZ does not penetrate
significantly into 
the He-shell, and consequently the intershell abundances are not
enhanced with carbon and oxygen, as has been found in models with
efficient PDCZ
overshooting. The mass fractions in the PDCZ of  the sixth TP are
($\hevi/\czw/\ose)=(0.854/0.143/0.002)$ (see below for the tenth TP).

In all successive thermal pulses deep DUP reaching below the
He-shell is present, although the mass dredged-up from below the
He-shell decreases somewhat (\abb{fig:mh1}). The events
during the tenth thermal pulse are shown in detail in
\abb{fig:HeCO-TPb}. The four abundance profile panels 
(i-iv) show the situation at the successive times indicated in the top
panel. Panel (i) shows the situation shortly before the onset of the
He-shell flash. The dip in the \ose\ profile at location (A) is a
signature of the H-shell. In the convective envelope (light shaded
region) \nvi\ is the most abundant species of the CNO elements, indicating
efficient HBB. At location (B) the large abundance of \nvi\
reveals the nature of this material as H-shell ashes. Toward
location (C) \czw, \ose, and as well \nezw\ gradually increase in
abundance, while \nvi\ decreases as a result of $\alpha$-capture
reactions. Location (C) harbors He-shell ashes which constitute
the top layer of the C/O core.

Shortly after $t=49.51\times 10^3\jahre$, the He-shell flash drives the
PDCZ. Panel (ii) shows the abundances at the peak extension of the
PDCZ (darker shading). Just above $m_r=0.9615\msun$  the tiny He-buffer
and the signatures of the H-shell can be seen. Thus, the PDCZ does not
reach into the H-rich envelope and no protons are ingested into the
convection zone. Such an occurrence has been suggested to be
an alternative site for the \spr\ in extremely metal
poor or zero-metallicity TP-AGB stars \citep{aoki:01}. If such a
mixing event occurs in intermediate mass stars it should be restricted
to metallicities lower than that studied here (Z=0.0001). 

During the pulse \nvi\ is synthesized into \nezw, and to some extent
into \mgfu. The latter is the result of the $\nezw(\alpha,\n)\mgfu$
reaction, which releases neutrons for the \spr. In addition, \czw\ and
\ose\ are now present in the intershell with a larger abundance than
before the He-flash ($(\hevi/\czw/\ose)=(0.782/0.198/0.017)$). The
\ose\ abundance now exceeds $1\%$ by 
mass, a value which is very similar to predictions of more metal-rich
models obtained without overshooting \citep{schoenberner:79}. We
discuss in \kap{sec:depov} how the abundance of this primary oxygen
depends on the overshooting. However, the presence of $1-2\%$ of
oxygen by mass in the intershell of AGB stars after a few initial TPs
should be considered as a \emph{minimum} at \emph{any}
metallicity. The third  
DUP will mix this primary oxygen into the envelope. In stars of
solar or only mildly sub-solar metallicity, the presence of primary
oxygen might be unnoticeable observationally, partly because of the
inherent difficulties in determining oxygen abundances, and partly
because the initial oxygen abundance is very close or even equal to
the oxygen abundance in the intershell. However, this obviously changes
 in TP-AGB stars of very low metallicity. If the initial
envelope abundance has extremely low CNO abundance, then the DUP
of material containing $X(\ose)=0.02$  should be noticeable. In
fact, the zero-metallicity models of both \citet{chieffi:01} and
\citet{siess:02} report a significant fraction  of oxygen in the
dredged-up material that is especially obvious in the less massive
cases without very efficient HBB. From 
the intershell abundances reported here, one can immediately make some
simple predictions. For example, if a hypothetical extremely metal poor
star has a C-overabundance of [C/Fe]=2.5, and if this
C-abundance comes from the heavy pollution by  a TP-AGB star that has
not experienced very efficient HBB (which would have destroyed some O) then the
prediction for oxygen would be $\mem{[O/Fe]}\ga 1.5$. This is so simply
because the predicted $X_C/X_O$ ratio in the dredged-up material is
$\la10$. 

In panel (iii) the situation during the third DUP following the
He-flash is shown. In location (D) the bottom of the envelope proceeds
inward, bringing to the surface \czw, \ose, \nezw, and other
species. This \nezw\ is 
the seed for \nadr\ production by HBB. 
In location (E), panel (iii) a small \mgfu\ pocket has formed from the
$\alpha$-capture reaction on \nezw, that releases neutrons for the
\spr. The neutrons released during the production of this \mgfu\
pocket are in addition to the neutrons from the \nezw\ neutron source
considered in the present \spr\ models. These neutrons are released during
a high-T radiative \nezw\ burning phase (along the dashed line in the
top panel of \abb{fig:HeCO-TPb}) between $t=49.515\times 10^{-3}\jahre$
and $t=49.53\times 10^{-3}\jahre$. The temperature during this
$15\jahre$ period of time decreases from the  peak PDCZ value of  $\log T=8.55$ to
$\log T=8.3$, and for $10$ out of the $15\jahre$, $\log T>8.4$. In
more metal rich or low mass TP-AGB models, the DUP  never
reaches this layer. Consequently, present \spr\ models do not need to
consider these neutrons, which usually are buried in the core. However, in
this massive TP-AGB model of very low metallicity, the
DUP picks this material up as it reaches below the
He-shell. This is shown in panel (iv), where the situation at the end
of the third DUP is shown. 

Two interesting things can be seen in panel (iv). Firstly,  because
the DUP now reaches below the He-shell, 
additional primary oxygen is brought to the envelope in
significant amounts. Thus, 
oxygen is dredged-up efficiently, despite the fact that practically very
inefficient overshooting has been assumed at the bottom of the
PDCZ.  Secondly, the \nvi\ abundance in the bottom region of the
convective envelope exceeds the C and O 
abundance. This means that H-burning already takes place at the bottom
of the envelope. HBB has already started and processes C and O
dredged-up from the core \emph{in situ}. 

This dredge-up is \emph{hot} as opposed to the situation in less
massive or more metal rich TP-AGB stars. In sequence E79-D4
the peak H-luminosity  during the efficient hot DUP events  is in the 
range $1.4 - 1.8\times 10^6\lsun$ and is not changing
in a systematic way from pulse to pulse. The large H-luminosity leads
to a peak in the surface luminosity (\abb{fig:rad-Lsurf}).
At low metallicity the envelope reaches
faster into deeper and hotter layers  than at higher metallicity. The
temperature in the 
overshooting layer during the DUP remains below the  value
obtained in the H-shell during subsequent HBB. However, the \czw\ abundance
is much higher in the intershell then in the H-shell. Therefore
H-burning is not limited by the slow p-capture of \nvi, but only by
the faster $\czw(\p,\gamma)\ndr$ reaction. As a result nuclear energy
production by H-burning in the overshooting layer during DUP is
several orders of magnitude larger then in the H-shell during HBB, and
in particular vigorous in the exponential overshooting
zone. However, even within the convective envelope the  H-burning luminosity
is still significant. In this  innermost region of the convective
envelope, the abundance of \czw\ and \nvi\ are determined by both
mixing as well as p-capture nucleosynthesis. Due to the coupled
treatment of time-dependent mixing and nucleosynthesis, the models
predict a continuous decrease from the large intershell values to the
small envelope abundance. In the transition zone at the very bottom of
the envelope convection, the H-burning luminosity decreases gradually  outward
from the peak value in the overshooting layer. For large core masses like for
the benchmark 5\msun\ case E79 this additional luminosity 
increases the DUP efficiency. Since the radiative
temperature gradient is proportional to the luminosity
($\nabla_\mathrm{rad} \sim l$)  efficient H-burning during the DUP
increases the convective instability, and models including this effect
show more efficient envelope enrichment at each TP. 
In \kap{sec:depov} the dependence of this effect on the overshoot
parameter is discussed.

The H-burning during the DUP has been noted before for Z=0 TP-AGB models with
a small amount of overshooting by \citet{chieffi:01}. Their H-burning 
luminosities are even similar. However, only one or two pulses with
this hot DUP have been followed due to the numerically demanding
nature of this phase.

In addition to the $Z=0.0001$ models, a comparison sequence with
$Z=10^{-5}$ has been computed. Overall the
properties are  
very similar. In particular, the deep DUP, the hydrogen burning during the
DUP phase, and the efficient \nvi\ production through HBB is
found in the  intermediate-mass stars of the lowest metallicities. In
the extremely metal 
poor cases, the envelope CNO abundance is dominated by the primary
component from the third DUP. It is this envelope CNO abundance
which determines to a large extent the evolution of envelope burning AGB
stars, including their yields. Since DUP is so efficient
and sets in very early, stars with even lower metallicity than
$Z=0.0001$ still behave in a very similar way.

\subsection{\spr\ considerations}
\label{sec:spr}
The main nuclear production site of the \spr\ elements is usually
associated with low-mass TP-AGB
stars \citep{gallino:97b,busso:99,goriely:00,busso:01a,lugaro:02a,herwig:02a}.
This, however, does not imply that massive AGB stars do not synthesize
the \spr\ elements at all. Very little is known theoretically about
the \spr\ at very low metallicity. IMS of this metallicity may have a
distinctive \spr\ signature  
that can be used as a diagnostic tool for the interpretation of
observations.  
Two reactions can release significant amounts of neutrons in
conditions encountered in TP-AGB stars. The $\cdr(\alpha,\n)\ose$
reaction is efficient for $\log T>7.95$ while the
$\nezw(\alpha,\n)\mgfu$ reactions requires $\log T>8.45$ to release
significant amounts of neutrons. 

According to current models of the
\spr\ in low-mass stars  \cdr\ releases neutrons under radiative
conditions during the interpulse phase and provides most of the
neutrons in the majority of  \spr\ enriched 
stars. It is also believed that this neutron source is responsible for
\spr\ isotopic signatures found in pre-solar meteoritic SiC
grains. For this mechanism, protons have to be partially mixed with
\czw\ at the intershell-envelope interface at or immediately after the
end of the third DUP. This partial mixing zone will heat up as
the intershell layer contracts to the pre-flash configuration. Then
\cdr\ forms which at a later phase during the interpulse will capture
$\alpha$-particles and release neutrons. This scenario  has been very
successfully applied in conjunction with stellar 
evolution models that do assume  an
ad-hoc H-profile for the partial mixing zone. As discussed in \kap{sec:ov} 
recent studies of  the effects of
rotation and/or overshooting  as a physical mechanism for the required
mixing have revealed a more 
complicated picture that is not well understood yet.  

In the very low metallicity IMS with  
deep DUP, a partial mixing zone forms at the end of the third DUP as a
result of the overshooting algorithm. The first difference compared
to the
low-mass models is that the \cdr\ pocket forms immediately. As
shown already in the previous section, the temperature at the bottom of
the convective envelope is high enough for H-burning during the actual
DUP period. The narrow \cdr\ pocket at the end of the third
DUP episode of the tenth TP (corresponding to panel (5) in
\abb{fig:HeCO-TPb}) is shown in \abb{fig:HeCO-TPc}. The \nvi\ pocket
which acts as a neutron poison is also evident. Since the DUP has
reached below the 
He-shell, the mole fraction of \hevi\ (for the subsequent activation of the
$\cdr(\alpha,\n)\ose$ reaction) is only a fraction of the
\cdr\ abundance in the pocket. For this reason, the \cdr\ neutron
source may be not or only inefficiently available in these massive
TP-AGB models. 

The $(\alpha,\n)$ reaction of \nezw\ releases neutrons in the
He-shell during the He-flash. This neutron source is activated very
efficiently. \nvi\ from which
\nezw\ is generated is the result of CNO cycling on previously
dredged-up primary C and O. The neutron source efficiency therefore depends 
on the peak temperature that is reached as well as the cumulative
efficiency of 
previous DUP episodes.  In this model the third DUP is very
efficient and therefore the \nvi\ abundance in the ashes of the
He-shell is rather large. This enhances the neutron release in the
PDCZ, and this effect is obviously more pronounced at lower
metallicity. A second contribution comes from radiative \nezw\ burning
already mentioned in \kap{sec:pulsetdup}. Following the peak
of the He-shell flash the temperature in the He-shell gradually
decreases. At this time the actual He-shell is no longer convectively
unstable. Neutron captures may still be possible, but now under
radiative conditions. Later this layer will be picked up by the third
DUP. This second
contribution will likely not exceed $10\%$ of the \spr\ synthesis in
the PDCZ, but may be important for details at certain
branchings. Finally, it may be worth considering the effect of \cdr\ 
left behind in the H-shell ashes. Since the primary CNO abundance can
be quite large due to efficient DUP, the ratio of \cdr\ to
\fese\ seed will be much more favorable than in more metal-rich
stars. On the other hand, due to the large \nvi/\cdr\ ratio in the H-shell
ashes, the efficiency of the trans-iron element production will be
limited. A quantitative analysis of the \spr\ in these models will be
presented elsewhere.

\subsection{Surface abundance evolution}
\subsubsection{H and He}
During the TP-AGB phase  the envelope abundance is altered by
the combined effect of DUP and envelope burning. After
14 thermal pulses  the
envelope hydrogen abundance of the E79-D4 sequence has  decreased from
initially $X(\mem{H})=0.665$ to 
$0.654$. Initially the H-abundance changes mainly because of the third
DUP. However, after the eighth thermal pulse, the
H-destruction due to HBB  dominates and during the following TPs the
H-depletion follows a roughly  linear decrease with time that can be
approximated by 
\begin{displaymath}
X(\mem{H})=-1.3\times10^{-4} \times t_3 + 0.6669 
\end{displaymath}
with $t_3=t/1000$.
Correspondingly the \hevi\ abundance increases from $X(\hevi)=0.335$
to $0.343$ and follows approximately the relation
\begin{displaymath}
X(\mem{\hevi})=1.15\times10^{-4} \times t_3 + 0.3327 .
\end{displaymath}

\subsubsection{C, O and primary N}
The combined CNO abundance increases steadily and follows the relation
\begin{displaymath}
X(\mem{C+N+O})=2.1\times10^{-5} \times t_3 + 4.0\times 10^{4} .
\end{displaymath}
Thus, at each thermal pulse, the mass fraction of CNO material in the
envelope increases by $\Delta X \approx 2\times 10 ^{-4}$.
The relative abundance of the CNO isotopes changes periodically with
the pulse phase (\abb{fig:abund1}). After the onset of efficient third
DUP, the temperature during the interpulse phase at the bottom of
the convective envelope (HBB temperature)
quickly settles in the range $9.1 - 9.6 \times 10^{7}\kelv$. Primary
C and O dredged-up from the intershell is cycled into \nvi.  Part of
that cycling occurs immediately during the DUP, which can be seen
from the increase of \cdr\ together with \czw\ and \ose\ at the
DUP time (\kap{sec:pulsetdup}). During this short H-burning DUP phase, the
\czw/\cdr\ ratio rises to values in the range $6 - 7$, which is much larger
than the CNO cycle equilibrium value of $\sim 3.5$ that is obtained
in the envelope in the second half of the interpulse cycle due to
HBB. 

Even though the HBB temperature  is 
high enough to burn oxygen, the effect of oxygen DUP is clearly
noticeable. In this model, third DUP of oxygen prevents an efficient
\ose\ depletion by HBB. The oxygen isotopic ratios evolve towards $\ose/\osi\ga
100$ during the interpulse phase, but larger values ($120 - 180$)
are present in the envelope during the earlier TPs, and in particular
during the H-burning DUP events. The $\ose/\oac \ga 2\times 10^6$ at
all times.  

The most important aspect of the CNO
abundance evolution is the significant primary \nvi\ production. After
the 14 TPs computed here without mass loss, the \nvi\ abundance is
twice the solar \nvi\ abundance. The envelope, that will be lost
eventually,  of this model star
contains at this point approximately $8\cdot10^{-3}\msun$ of primary \nvi.

\subsubsection{Oxygen and Sodium}
\label{sec:oxysod}
The elemental abundance evolution of sodium and oxygen is correlated
or stationary in all models reported here. Oxygen in the envelope
increases because of dredged-up, and decreases because of HBB. In this
model these two effects balance each other. Sodium forms early on
in the pulse cycle, and is then destroyed again. This concordance can
easily be explained. \nezw\ -- the seed for \nadr\ -- follows indirectly
the dredge-up of primary C and O from the intershell, as it is made
from the \nvi\ that results from H-shell burning of the dredged-up C and
O material. If the envelope burning temperature is large enough for
the $\nezw(\p,\gamma)\nadr$ reaction, then the production of \ose\ and
\nadr\ are correlated. The 
destruction, on the other hand, is also correlated because the p-capture
rates of both isotopes are very similar \citep{angulo:99}. 

\citet{denissenkov:03a} have discussed the implications of these
processes in massive AGB stars for the O-Na anti-correlation
in globular clusters. They have argued that low-metallicity
IMS  are probably not directly responsible for 
the observed  O-Na 
anti-correlation. An important
detail in their argument is the recurrent DUP of oxygen
(\kap{sec:pulsetdup}), which 
would require larger HBB temperatures for effective oxygen
depletion. At such large temperatures any sodium produced by p-capture
on \nezw\ would be destroyed as well. In a star with lower HBB
temperatures, simultaneous oxygen and sodium enhancement is
predicted. In any  case,  AGB stellar models predict an O-Na correlation,
or no correlation, rather than an anti-correlation.
In \abb{fig:O-Na_3} the sodium and oxygen abundances the globular  
cluster stars and the model prediction from this work are
compared. The globular cluster
stars show a varying degree of oxygen depletion that is associated
with a sodium enhancement. 

The offset in sodium abundance between the model abundances and the
 globular cluster stars can be fully attributed to the choice of
 initial composition for the models.
The TP-AGB initial sodium abundance of the models, which correspond to
 $\mem{[Na/Fe]}\sim 0.6$ and 
 $\mem{[O/Fe]}\sim -0.2$  (\abb{fig:O-Na_3}), are the result of the
 2$^\mem{nd}$DUP, during which the envelope convection reaches into
 the ashes left behind by the H-shell. In the H-shell, primordial 
\nezw\ is partially transformed into \nadr. The initial sodium
abundance of the TP-AGB model corresponds to an initial scaled-solar 
 \nezw\ abundance. While no observational information on the noble gas
neon is available for low metallicities, theoretical models of 
galactic chemical evolution agree that  
[Ne/Fe] is larger at lower [Fe/H].  At [Fe/H]$=-2.5$ the predictions
range from [Ne/Fe]$=+0.2$ \citep{timmes:95} to [Ne/Fe]$=+0.7$
\citep{alibes:01}. However, this is predominantly \nezwa, since  \nezw\ is
 a secondary isotope and follows the initial abundance of \nvi\ in
 massive stars \citep{woosley:95}. Unless the parent cloud of the AGB
stars modeled here 
has been polluted by an even earlier generation of massive AGB stars,
 its \nezw\ 
 abundance has probably been negligible. 
In that case, the initial sodium 
abundance of the Z=0.0001 massive TP-AGB stellar models is the same as
 the initial sodium abundance at the time of star formation.  In order
to explain the O-Na anti-corelation in globular clusters by
simultaneous Na production and O destruction it is
often assumed that $\mem{[Na/Fe]}\lesssim 0$  initially at low metallicity.  
This is in agreement with theoretical models of the galactic chemical
 evolution  \citep{timmes:95,alibes:01}, although a systematically lower
 [Na/Fe]  at lower [Fe/H] is not obvious from the analysis of field
 giants by \citet{gratton:00}. However, they report a scatter of
 $\sim \pm 0.3$ around $\mem{[Na/Fe]}=0$ for $-2.5<[Fe/H]<-0.5$.

\subsubsection{Magnesium}
The magnesium isotopic ratios can be modified by 
three processes in AGB stars: (a) $\alpha$-captures on \nezw\ in the
He-shell (flash), which can produce both \mgfu\ and \mgse, (b) $\p$-captures
on \mgvi\ and \mgfu\ during the envelope burning phase, and (c)
by n-captures on  \mgvi\ and \mgfu\ as well as on \alse. In
\citet{denissenkov:03a} the  TP-AGB evolution dominated by high HBB
temperature is discussed, during which the initially available \mgvi\
burns mainly into \mgfu. In cases of lower HBB temperature (either
because of the inclusion of mass loss or because of a smaller mass),
and if a complete treatment of neutron capture reactions is taken into
account, the other two effects operating in the PDCZ become more
important for the model predictions.

\subsection{Dependence on the overshooting parameter}
\label{sec:depov}
Three additional tracks with different assumptions on the overshooting
efficiency have been calculated (\tab{tab:ovdep}). Comparison of
sequence D4 and D9 
reveals the influence of the overshooting parameter at the bottom of
the PDCZ. For sequence D9 a larger overshooting  efficiency
($f_\mem{PDCZ}=0.016$)  than for
sequence D4 has been assumed. These tests confirm the  trends found
for the  3\msun, Z=0.02 stellar model with intershell
overshooting by \citet{herwig:99a}. 
In comparison to the D4 sequence, the D9 sequence shows a
\begin{enumerate}
\item larger  $L_\mem{He}$-flash peak  luminosity by roughly a
factor $1.5 - 2$
\item longer interpulse period by $\sim 20\%$
\item slightly smaller DUP H-burning luminosity, and similar
quiescent interpulse H-shell luminosity
\item larger third DUP, e.g.\ at the sixth TP of the D9 sequence
at $t-t_0=35384\jahre$ ($t_0$ as in \abb{fig:mh1}) the
DUP parameter is $\lambda=1.46$ compared to $\lambda=1.18$ at the
eighth TP of the D4 sequence at $t-t_0=32644\jahre$; cumulatively this deeper
DUP leads to a somewhat stronger core mass reduction than in the
D4 case; the core mass at the end of the sixth TP of the D9 sequence
is $m_r=0.95539\msun$, and
\item a different He/C/O intershell abundance composition; C
and O are enhanced at the expense of He; in the PDCZ of TP6 the mass
fractions are (He/C/O)=(0.62/0.26/0.18).
\end{enumerate}
Despite these differences, overshooting at the bottom of the PDCZ does
not change the envelope abundance evolution qualitatively.
In both cases, with and without overshooting at the bottom of the PDCZ,
the DUP is efficient, and substantial amounts of primary C and O
will be dredged-up. Sequences of intermediate-mass and
massive AGB models generally tend to converge to $\lambda=1$ whether
they start initially with  $\lambda>1$ or $\lambda<1$
\citep{karakas:02a}. This phenomenon 
is already noticeable in \abb{fig:mh1}. Therefore, the difference in DUP
efficiency is expected to decrease in later pulses. The 
different \ose\ and \czw\ abundances in the PDCZ lead to some
differences in the surface evolution of the CNO elements. For a
comparison, the abundance evolution of the D4 and D9 case are shown in
 \abb{fig:abund3} in the logaritmic square bracket units used to present
abundance observations. Deeper DUP
and a larger \czw-abundance in the PDCZ cause a larger enrichment in primary
\nvi\ in the D9 case compared to the D4 case. A difference
of the \ose\ abundance of a few tenths of a dex is present. If many
more TPs occur in this type of stars, the difference of the oxygen
abundance in the two cases would likely increase. However, overall the
uncertainty in the model predictions of massive AGB stars do not
appear to be strongly  dependent on the overshooting efficiency at the
bottom of the PDCZ. 
\begin{deluxetable}{lllll} 
\tablecolumns{5} 
\tablecaption{\label{tab:ovdep}
Model sequences used in this paper} 
\tablehead{ 
\colhead{track ID} & \colhead{$f_\mem{PDCZ}$}   &
\colhead{$f_\mem{CE}$}& \colhead{$M_\mem{ZAMS}/\msun$}
& \colhead{$Z$}}
\startdata 
E79-D4$^a$ & $0.002$ & $0.016$ & $5$&$0.0001$\\ 
E79-D9 & $0.016$ & $0.016$& $5$&$0.0001$\\ 
E79-D10,11& $0.002$ & $\geq 0.032$& $5$&$0.0001$\\ 
E37-D2 & $0.016$ & $0.016$& $5$&$0.02$ \\ 
E85-D2 & $0.016$ & $0.008$& $4$&$0.0001$\\ 
\enddata 
\\
$^a$benchmark sequence
\end{deluxetable}

Next, the influence of the exponential overshooting efficiency at the
bottom of the convective envelope during the DUP phase is
analyzed. A number of tests (among them E79-D10 and D11 with
$f_\mem{CE} \ge 0.032$) have been followed for several ten thousand
models with small time steps during the hot DUP phase. In some cases the mixing speed
has been altered too. The dredge-up
rate was almost constant  in all cases ($\mem{d}m_\mem{bce}
/\mem{d}t \simeq 5\cdot10^{-5}\msun/\jahre$ for E79-D10, with
$m_\mem{bce}$ is the 
Lagrange  coordinate of the bottom of the convective envelope), and depends
somewhat on both the assumed speed and  
depth of the exponential  overshooting.
The surface abundance evolution during the ongoing hot DUP after the fifth TP
(e.g. sequence E79-D10 which has been followed longest,
\abb{fig:cno-obsunits10}) is similar to the long-term evolution over
many TPs with moderate DUP (\abb{fig:abund3}). However,
in the latter case the N/O and the N/C ratios are larger and the C/O
ratio is smaller.

When the sequence
was stopped the dredged-up mass was $\Delta M_\mem{DUP}=1.6\cdot
10^{-2}\msun$, which is about eight times the mass contained in the
He-shell flash convection zone. During the hot DUP the stellar
luminosity is about $\log L/\mem{L_\odot} = 5$, again somewhat
dependent on the details of the overshooting. With the large 
envelope CNO abundance enhancement the mass loss may not be  less
efficient than in more 
metal-rich AGB stars. Assuming the mass loss rate of \mbox{Bl\"ocker
(1995)} with an efficiency parameter $\eta_\mem{BL} = 0.1$ the mass loss
during this phase is of the order $\mdot=2\cdot 10^{-3}\msun/\jahre$.
Together with an envelope mass of about $4\msun$ this implies an upper
limit for this evolution phase of $t_\mem{DUP} < 2000\jahre$, and with
the  dredge-up rate the maximum mass of dredged-up material
is $\Delta M_{DUP} \sim 0.1\msun$.

It is emphasized that these findings are the result of \emph{test}
calculations intended to explore the uncertainties of the model
predictions. Based on
the current models, and without good constraints on overshooting in the
presence of nuclear burning,  we 
can not exclude that IMS of very low metallicity end their lives in a
 high mass loss hot-DUP phase, possibly after a rather early
thermal pulse.

\section{Conclusion}
\label{sec:concl}
The evolution of massive AGB stars (mainly of initial mass
$5\msun$  and  very low metallicity, $Z=0.0001$) has been studied in
detail. Nucleosynthesis and 
convective mixing, including the effect of exponential overshooting
has been considered. In contrast to less massive cases and more metal
rich cases this model predicts hot DUP, the efficient nuclear burning
of hydrogen during the third DUP phase which can greatly increase the
DUP efficiency. The additionally released energy feeds
back into the structure evolution, leading to deeper
DUP  than observed in models that do not show this effect
(either because strictly no mixing beyond the Schwarzschild boundary
of convection is allowed or because the mass and metallicity are not in the
appropriate range). In the  benchmark case a DUP penetration through
and beneath the He-shell is found even with a rather small
overshooting efficiency. For larger masses or
larger overshoot efficiencies, the DUP is deeper and may eventually
terminate the AGB evolution.

Our model predictions of O and Na in massive low-Z TP-AGB models is
not in agreement with the trend seen in the globular clusters, and
this casts doubt on the simple self-pollution scenario of globular
clusters. 
On the other hand AGB stars have been frequently associated with the
significant 
overabundance of C and N in some EMP stars. 
 EMP binaries CS\,29497-030 \citep{sivarani:03} and
LP\,625-44 \citep{aoki:02} show oxygen and sodium enhancements similar
in magnitude to the model predictions presented here. However, the
majority of C-rich EMP (including these two) stars have a larger C
than N enhancement. Such an abundance 
pattern is incompatible with HBB in massive AGB stars, in which at all
metallicities C and O is cycled efficiently into N (\abb{fig:abund3} and
\ref{fig:cno-obsunits10}). 

The   EMP star HE\,0024-2523 \citep{lucatello:02} is a
post-common envelope binary with a period of only a few days. Our
models suggest that the most likely time to initiate the common
envelope evolution should be at the time of the first thermal pulse
with efficient DUP. In the 5\msun\ sequence this is the fifth
TP. During this phase the star expands temporarily very significantly
(\abb{fig:rad-Lsurf}).  The radius peaks occur at all subsequent 
thermal pulses, but the maximum radius during these peaks is not
significantly exceeding the initial radius peak. Thus, if the common
envelope evolution has not been initiated during the preceeding  RGB
or early AGB evolution, the most likely time is at the first TP with
efficient DUP. We do not model the common envelope evolution in this
paper. However, it is interesting to note that the C/O ratio in the
intershell is about 100 during the first TP with deep DUP
\kap{sec:pulsetdup}. During the following TPs the intershell C/O ratio
decreases to $\sim 15$ at the tenth TP.  With this C/O ratio in the
intershell abundance the observed C/O ratio in HE\,0024-2523 of 100
can not be reproduced no matter how peculiar the  burning and mixing 
induced by the common-envelope evolution may be. However,
HE\,0024-2523 is highly \spr\ enhanced and it is not clear how this
can be explained without a succession of neutron exposures in many
subsequent TP cycles. This example shows that the \spr\ in AGB stars
of very low metallicity is not very well understood yet.

Our calculations confirm the previously found trend that both a lower
metallicity as well as a larger core mass increase the DUP
efficiency in terms of the DUP parameter $\lambda$
\citep[e.g.][]{boothroyd:88}. \citet{marigo:01b} and \citet{marigo:99} found
semi-empirically that  
low-mass and moderately metal-deficient AGB stars should already have
a substantial DUP parameter of $\lambda \sim 0.5$. The fact that
our models show larger $\lambda$ is generally consistent with this trend.
It is shown that efficient dredge-up can be found even under the
assumption of small overshooting efficiency, and that a high numerical
resolution is required for dredge-up predictions.

\acknowledgments
I am indebted to  D.\,A.\,VandenBerg for his encouragement and
generous support through his
Operating Grant from the Natural Science and Engineering Research
Council of Canada. I would also like to thank T.\,Beers, P.\ Denissenkov
and J.\,Johnson for many interesting discussions on abundances
in stars of very low metallicity. D.\,Sch\"onberner and N.\,Christlieb
have kindly helped me by critically reading an earlier version of the
manuiscript.


\clearpage
\twocolumn
\begin{figure}
\plotone{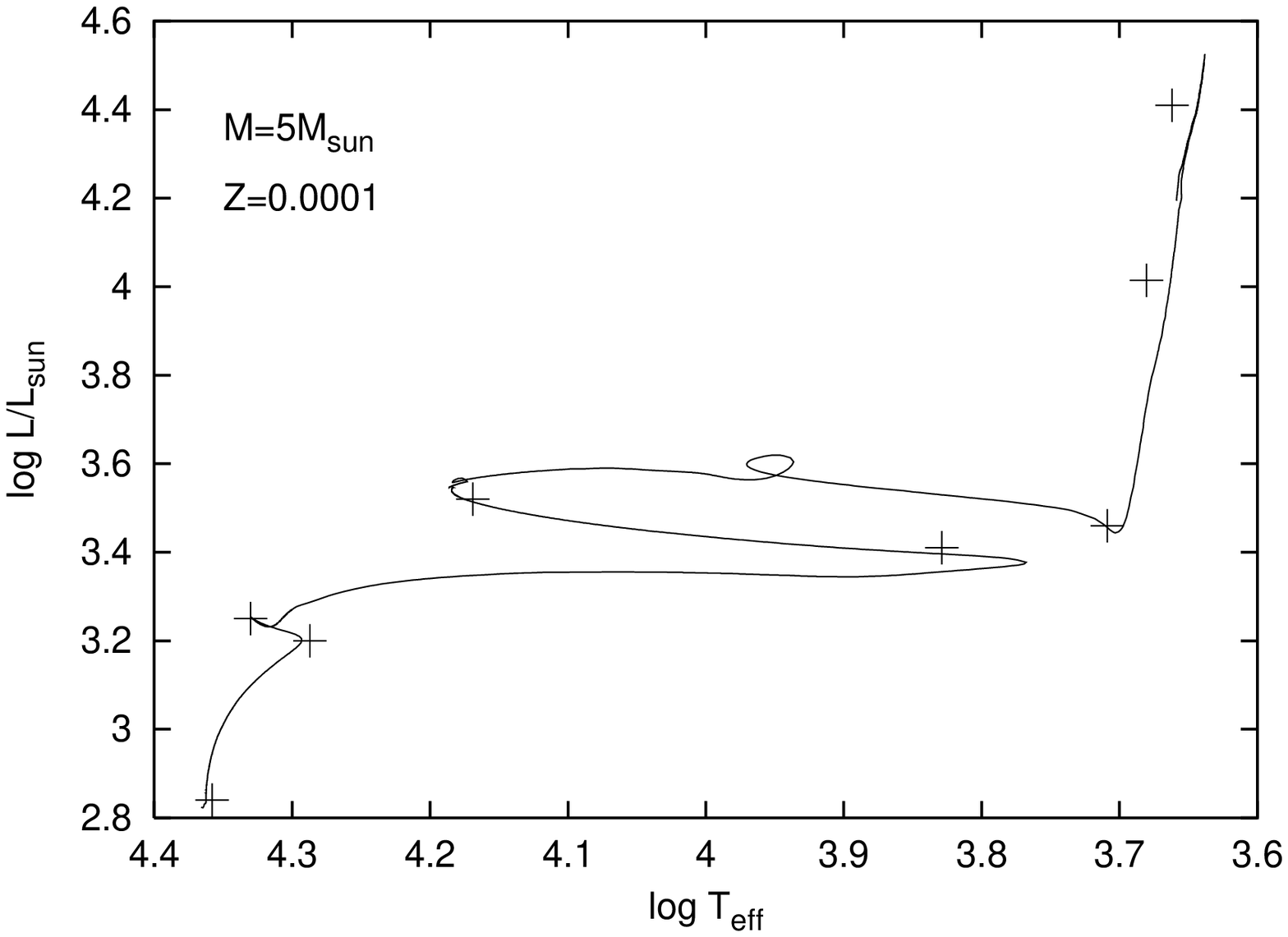} 
\figcaption{ \label{fig:hrd-preAG} 
HRD of pre-AGB evolution. The plus signs along the track indicate 
selected points along an evolution sequence computed  by
\citet{girardi:96} with the same initial mass and metallicity.
}\end{figure}
\begin{figure}
\plotone{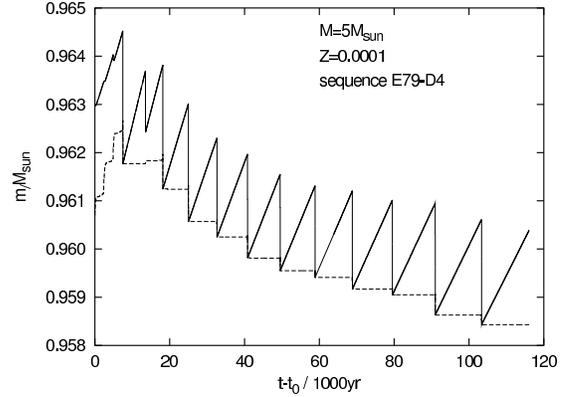} 
\figcaption{ \label{fig:mh1} 
Evolution of the hydrogen-free core (solid line, defined as
$\mh=m_r(X_\mem{H}<0.37)$) 
and  the helium-free core (dashed line, defined as
$\mhe=m_r(X_\mem{He}<0.49)$) during 
the TP-AGB. $t_0=99385139.9\jahre$ at maximum of $L_\mem{He}$ during
first thermal pulse. $M=5\msun$, Z=0.0001, sequence E79-D4.  See text
for an explanation of the small DUP at $t-t_0 \sim 15\cdot 10^3\jahre$.
}\end{figure}
\begin{figure}
\plotone{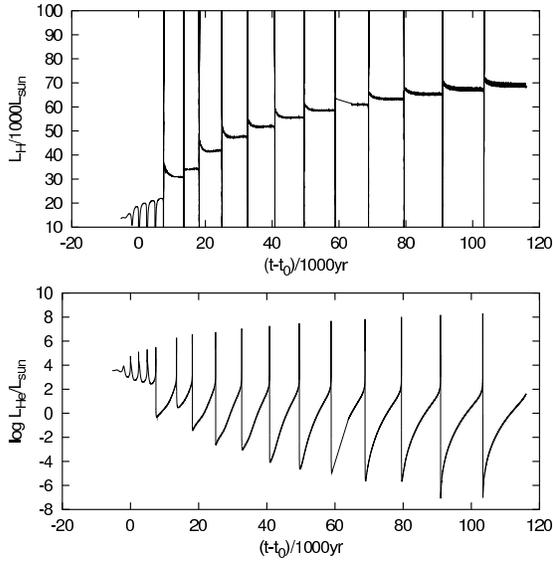}
\figcaption{ \label{fig:lumi} 
Evolution of the hydrogen-burning luminosity (top panel) and the helium
burning luminosity (bottom panel) for the same sequence as shown in \abb{fig:mh1}.
}\end{figure}
\begin{figure}
\plotone{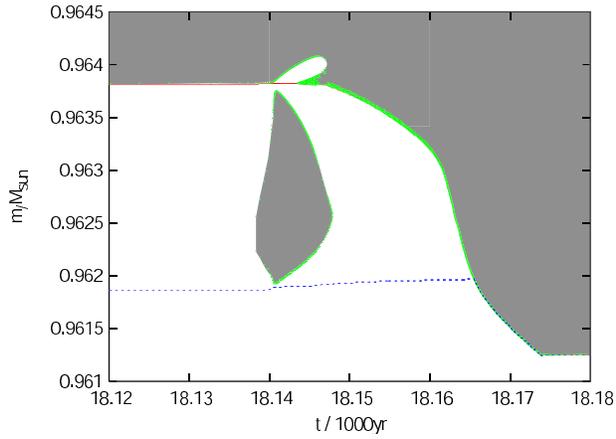}
\figcaption{ \label{fig:t-mh1-tp6} 
Evolution of the hydrogen-free core, the helium-free core,  and the convection zones during the sixth
TP ( $M=5\msun$, Z=0.0001, sequence E79-D4). Except for a short period immediately 
after the He-flash at $t=18.14\cdot10^3\jahre$, the bottom of the
envelope convection zone and the H-free core coincide in this
representation. Note that the third DUP penetrates into the
He-free core. $t_0=99385139.9\jahre$ at maximum of
$L_\mem{He}$ during first thermal pulse.   
}\end{figure}
\begin{figure}
\plotone{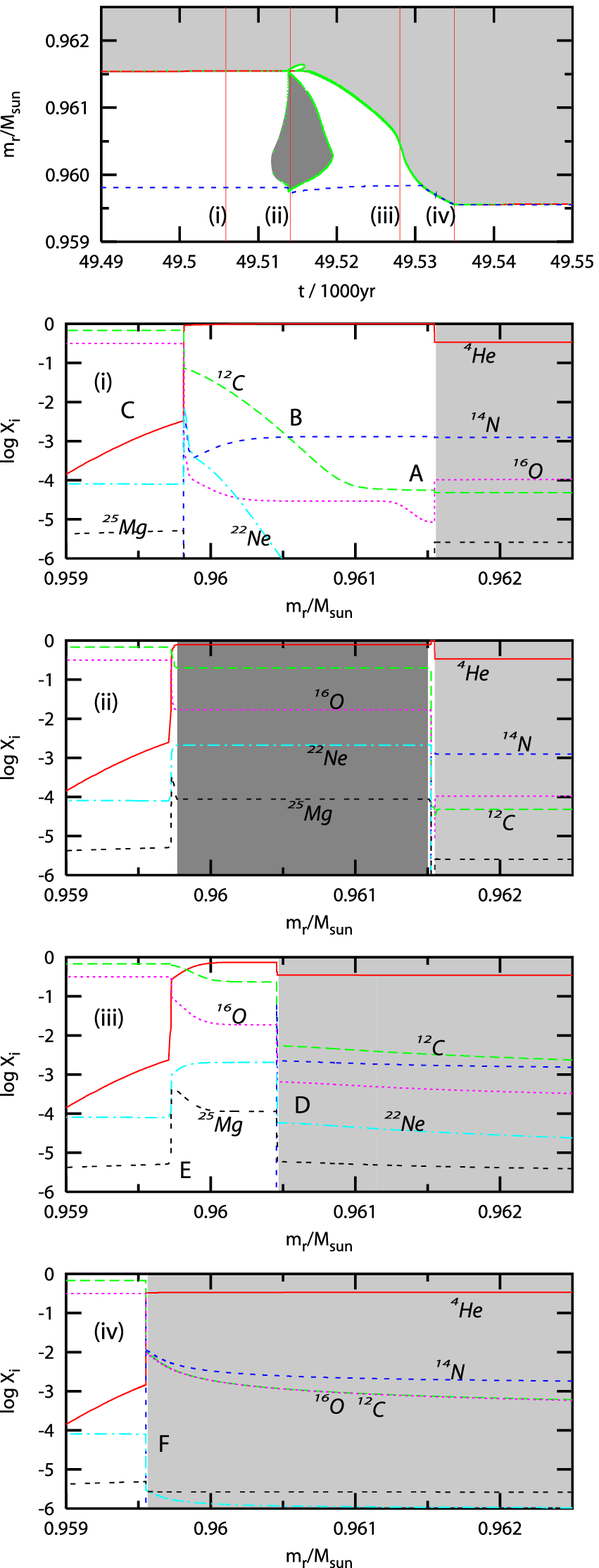}
\figcaption{ \label{fig:HeCO-TPb} 
for caption see page \pageref{cap:HeCO-TPb}}
\end{figure}
\newpage
\label{cap:HeCO-TPb}
Fig.\ \ref{fig:HeCO-TPb} -- {Panel (1): Evolution of the Lagrangian
location of the H- and He-burning shells and the 
convectively unstable layers (dark shade = PDCZ, light shade =
envelope convection). Panel (2-5): Abundance profiles at times
indicated by thin vertical lines in panel(1). Labelled regions of
special interest are discussed in the text. A blow-up of region F in panel
(iv) is shown in \abb{fig:HeCO-TPc}.
\newpage

\begin{figure}
\plotone{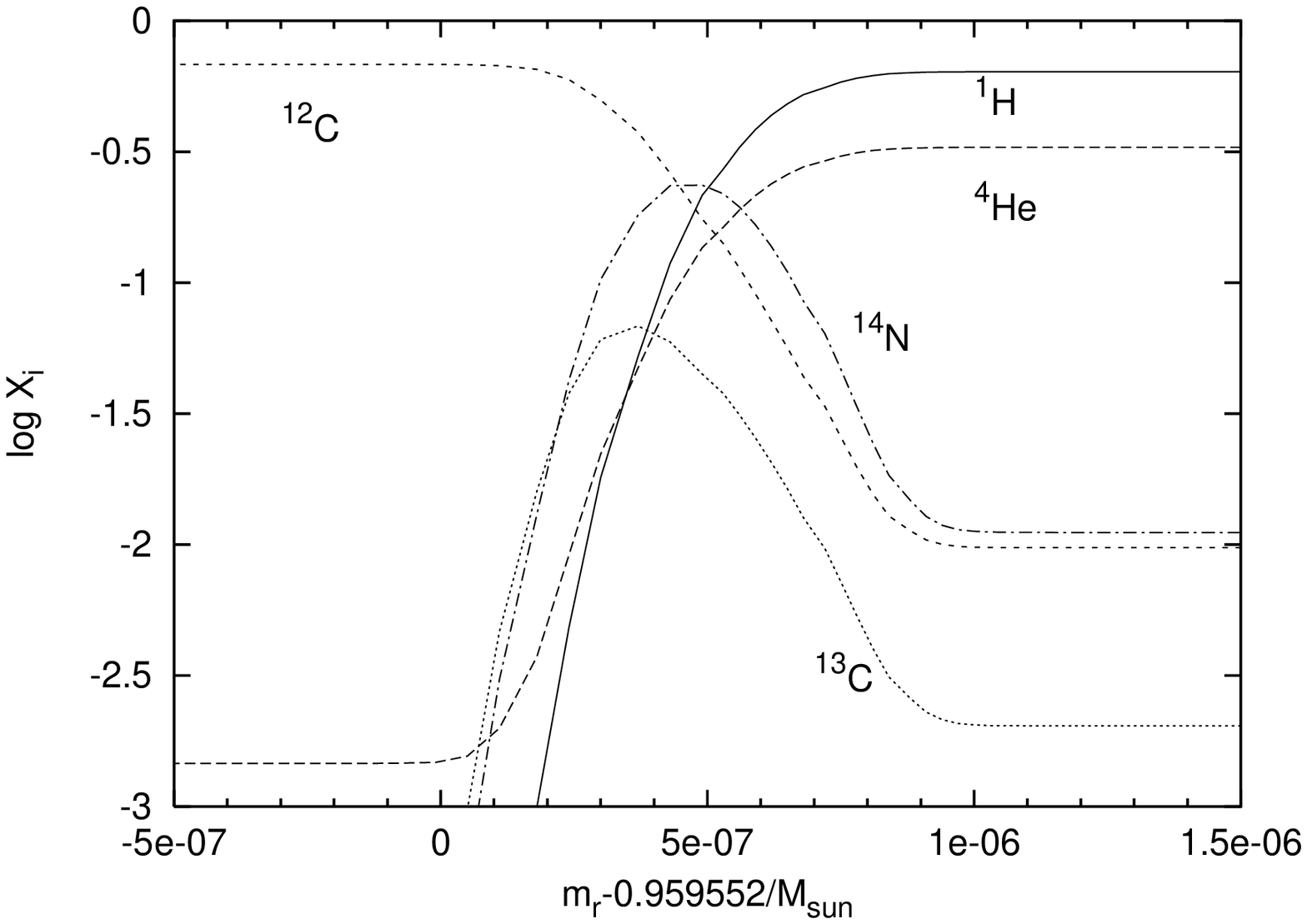}
\figcaption{ \label{fig:HeCO-TPc} 
Abundance profiles of species relevant for the \spr\ in region F of
\abb{fig:HeCO-TPb}, panel (iv). The region on the left is the top-layer
of the C/O core, the region on the right is the bottom of the envelope
convection zone at the end of the third DUP after the tenth
TP. The helium abundance in the layer below the convection zone is
low because the third DUP has reached below the He-shell (see text). 
}\end{figure}
\begin{figure}
\plotone{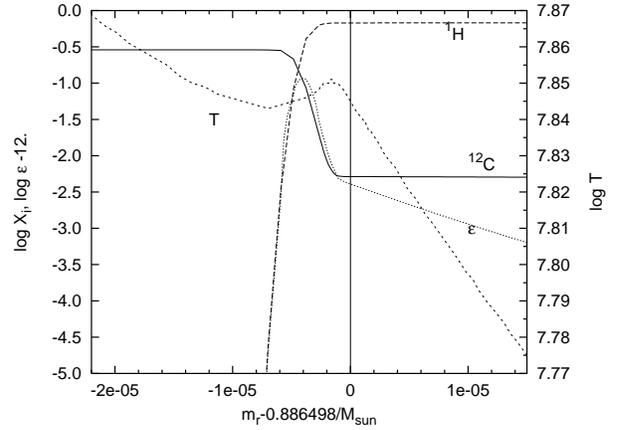}
\figcaption{ \label{fig:E85-Hburnprofile} Profiles of protons, \czw,
nuclear energy generation $\epsilon$ (left scale), and temperature
(right scale) at the bottom of the convective envelope during the
DUP of sequence E85 ($Z=0.0001$). The
abscissa has been shifted by the mass coordinate of the envelope
convection boundary, and the region to the right of the vertical line is the
 convective envelope.}
\end{figure}
\begin{figure}
\plotone{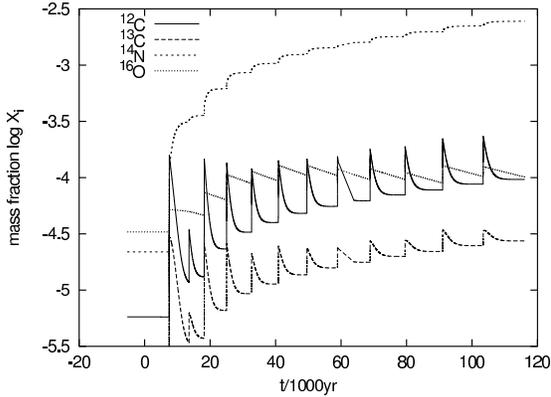}
\figcaption{ \label{fig:abund1} Envelope abundance evolution of
relevant CNO isotopes for the 5\msun\ benchmark case (E79-D4). Two irregular
behaviours can be seen. At $t\sim 15\cdot 10^{3}\jahre$, the DUP
is smaller than after the preceding and the following thermal pulse
as the result of insufficient numerical resolution (see
\kap{sec:pulsetdup} for details). At $t\sim 62\cdot 10^{3}\jahre$, some
data files for the plot routine were lost, and the  HBB phase
following the thermal pulse -- most notably of \czw\ -- is not plotted
accurately.  
}\end{figure}
\begin{figure}
\plotone{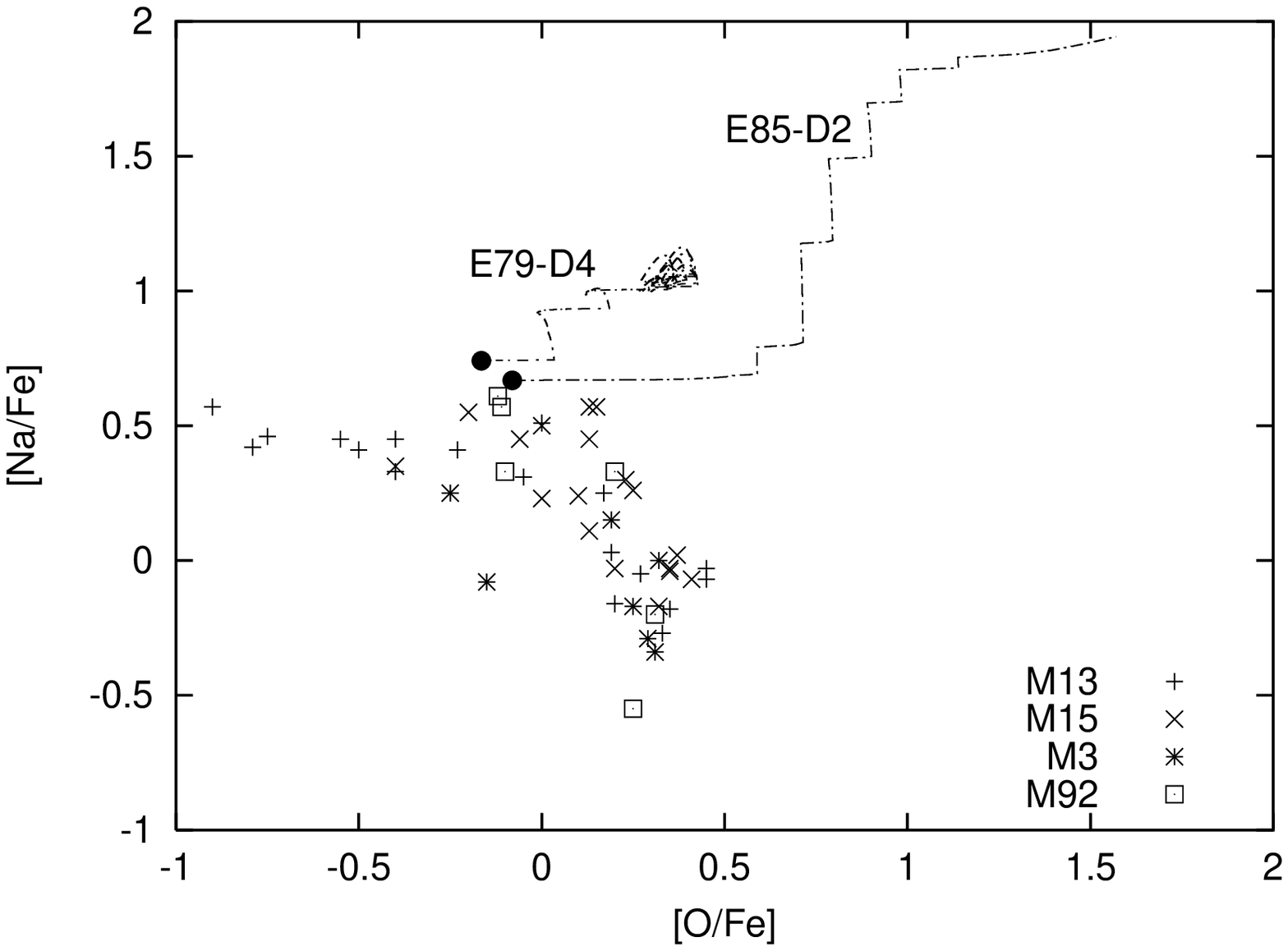}
\figcaption{ \label{fig:O-Na_3} 
Oxygen and sodium abundances in globular cluster stars
\citep{langer:97,sneden:97,shetrone:96,kraft:92} and  AGB model
predictions (\tab{tab:ovdep}). The $5\msun$ 
sequence E79-D4 has been computed without mass loss and achieves high
HBB temperatures, while the $4\msun$ sequence E85-D2 has been computed
with mass loss and neither oxygen nor sodium are destroyed by
HBB. Both sequences start with low O and Na abundances indicated by a
filled circle (see text for details). For the conversion of the model
abundance mass
fractions into the logarithmic abundance ratios with respect to solar
$\log \ose_\odot=-2.018$, $\log \nadr_\odot=-4.476$ and  $\log
\fese/{\fese_\odot}=-2.3$ 
has been used. 
}\end{figure}
\begin{figure}
\plotone{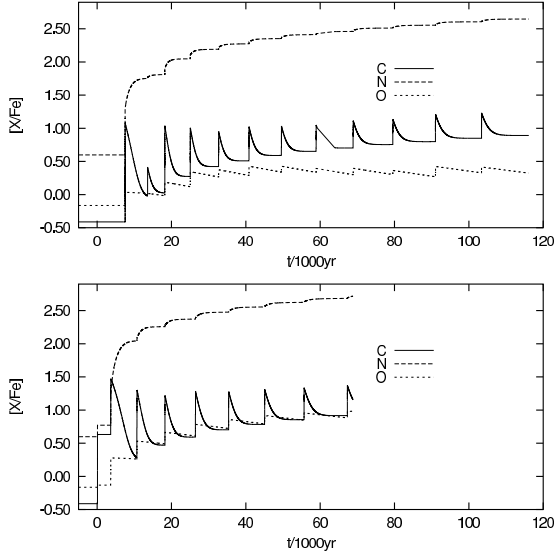}
\figcaption{ \label{fig:abund3} 
Abundance evolution in units of
[X/Fe]=log(X/Fe)-log(X$_\odot$/Fe$_\odot$), with $\log
\ose_\odot=-2.018$, $\log \czw_\odot=-2.479$, $\log \nvi_\odot=-2.957$
and  $\log \fese/{\fese_\odot}=-2.3$. Top panel: sequence D4
($f_\mem{PDCZ}=0.002$, $f_\mem{CE}=0.016$); bottom panel: case D9
($f_\mem{PDCZ}=f_\mem{CE}=0.016$) (see \tab{tab:ovdep}). 
}\end{figure}
\begin{figure}
\plotone{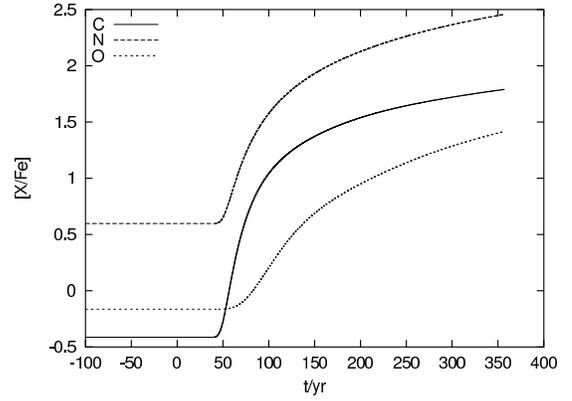}
\figcaption{ \label{fig:cno-obsunits10} 
Abundance evolution in the same units as in \abb{fig:abund3} for  
test calculation E79-D10 after the fifth TP (see text for details). $t=0\jahre$ at maximum 
$L_\mem{He}$ during the flash.
}\end{figure}

\end{document}